# Pulsed high-power radio frequency energy can cause non-thermal harmful effects on the brain

Omid Yaghmazadeh

*Abstract*— High-power microwave applications are growing for both military and civil purposes, yet they can induce brain-related risks and raise important public health concerns. High-power sub-millisecond radio frequency energy pulses have been demonstrated to be able to induce neurological and neuropathological changes in the brain while being compliant with current regulatory guidelines' limits, highlighting the necessity of revising them.

## I. Introduction

Ever since the primary applications of Radio Frequency (RF) energy waves, their interactions with the biological tissue have been of immense interest for potential health concerns making them an important topic in public health [1], [2]. Novel advances in the development of high power microwave (HPM) applications, providing the possibility of producing very short time-scale extremely high-power RF pulses, has brought new insights in this subject and raised novel concerns regarding RF energy radiation exposure. Some recent studies have shown that application of short high power RF pulses can lead to brain damage with potential severe impact. This raises an important concern in environmental health as there is a growing trend for HPM applications. Such health effects should therefore be studied more carefully and regulatory guidelines should be updated to avoid their impact on the society.

## II. Non-thermal Effects from the Biophysical vs. the Biological Perspectives

The brain, due to its electrical nature, has long been considered the most vulnerable organ to exposure to Radio Frequency (RF) energy waves. It is well established that RF energy can cause heating of the biological tissue, such as the brain, and consequently affect its function (as it is established that local increase in brain temperature affects ongoing neural activity [3]). There has been, however, a long-lasting question whether RF energy is able to affect neuronal activity and brain function in a non-thermal manner. This question has been at the center of a controversial debate where theories and experiments both in favor and against such possibility have been reported over several decades [4]. The non-thermal effects of RF energy on the biological parameters can be considered from two different angles: 1) form the "biophysical" point of view, non-thermal effects of RF energy are such effects whose mechanisms are not thermal (nor thermally-mediated); 2) from the "biological" point of view, non-thermal effects of RF energy are any effects that do not increase the tissue temperature in a biologically meaningful timescale. In this context, if a stimulus, such as short high power RF pulses, is capable of affecting the tissue, by a thermal (or a thermally-mediated) mechanism without significantly increasing the overall tissue temperature, their effects are considered non-thermal from the biological point of view but not from the biophysical point of view.

## III. RF Exposure Can Affect (and Damage) the Brain in a Biologically 'Non-thermal' Manner

Recently, Hao et al. reported an experimental demonstration of biologically non-thermal effects of RF energy on the brain of freely behaving mice that disturbed brain cognitive functions related to learning and memory [5]. They used RF energy at 2.856 GHz, pulsed at 80Hz with 0.5ms pulse width and a power density of 200 mW/cm$^2$. This exposure didn't induce changes in the mice rectal temperature beyond the biological range (<1C). They examined various hippocampus dependent spatial learning and memory tasks and reported significant changes in animal's performance/behavior for up to several days. They also found that the release of dopamine was significantly reduced in the CA1 hippocampal region of the exposed mice for several days and reported that the axonal projection from locus coeruleus dopaminergic neurons to dorsal hippocampus were weakened upon RF exposure. In addition, they observed abnormalities in the structure and molecular mechanism involved in the dopamine synapses in the hippocampal region of the RF exposed mice.

In another recent study, Dagro et al. studied the thermos-elastic effects of short duration high-power RF pulses on affecting the human brain using a computational approach [6].

This manuscript was submitted on August 16$^{th}$, 2023. OY is supported by the NIH-TL1 postdoctoral fellowship grant (#2TL1TR001447-06A1) and partially by the NIH-R01 (# 1R01NS113782-01A1) grant.

O. Yaghmazadeh is with the Neuroscience Institute, School of Medicine, New York University, New York, NY 10016, USA (correspondence e-mail: omid.yaghmazadeh@gmail.com).

Their results demonstrated that short-timescale high-power single RF pulses are able to induce transient and fast thermal expansion in the brain that can cause mechanical stress on the brain tissue leading to neurological effects. Particularly, they show that for very short (few μs) and extremely high-power (with an incident power density of $1\times10^7$ mW/cm$^2$) single RF pulses, the induced mechanical stress could exceed injury threshold and cause permanent neuropathological damage while the applied RF exposure is biologically non-thermal and compliant with regulatory limits [6].

Although the required field strength to induce injuries as studied in Drago et al.'s report is extremely high (yet still accessible with current HPM developments for military, civil or research applications), their simulations also demonstrated the induction of a high level of mechanical stress inside the brain even for lower power levels and longer pulses. In addition, application of repeated pulses, which is not studied in their report, could have accumulating effects on the brain. This is indeed demonstrated in Hao et al.'s study where application of a repeated pulse sequence with larger duration (500μs) and much lower power density (200 mW/cm$^2$) lead to neurological effects that last over, at least, several days. Although their histological evaluation did not show any lesion, they reported some structural deformation related to the dopamine-related mechanisms. This indicates that repetitive sub-millisecond pulses at relatively (but not extremely) high power levels can induce brain damage and cognitive deficits in a biologically non-thermal manner.

Lastly, in a third study, Yaghmazadeh et al. examined effects of continuous-wave (CW) RF energy exposure on the neural activity in a head-fixed mouse set-up in-vivo [7]. Using 1-photon Ca$^{2+}$ imaging in mice brain, they reported that RF energy radiation, up to power levels that induce a local point SAR value of 28.8 W/Kg, does not alter the neuronal activity with statistical significance. Their study confirmed that for a relatively high level of CW RF energy radiation (several fold higher than what is permitted by the regulatory limits [7]) neuronal activity is not affected in a non-thermal manner (from both biophysical and biological point of view).

## IV. DIFFERENT EFFECTS OF PULSED VS. CW RF ENERGY ON NEURAL ACTIVITY

Hao et al. compared their findings with the results in Yaghmazadeh et al.'s report and stated a contradiction in the outcomes: while the latter report stated absence of non-thermal effects of RF energy exposure on neuronal activity, their results demonstrated a RF exposure paradigm that led to significant non-thermal effects. They attributed the contrariety of these findings to the difference in the applied frequency (2856Hz in their study against 950MHz in the other). However, there is no demonstrated mechanism that could explain why neurons would respond differently to distinct frequencies at the GHz range. Moreover, besides the applied frequencies, the major difference in the experimental paradigms that were used in these two studies lies under the fact that Hao et al. used sub-millisecond pulsed RF energy (with several fold higher power density) while Yaghmazadeh et al. used continuous-wave stimuli. Consequently, while Hao et al.'s results demonstrate a 'biologically' non-thermal mechanism for affecting the brain, its 'biophysical' mechanism is indeed thermally mediated (thermo-elastic waves in the tissue are in fact initiated by thermal energy deposition). On the other hand, Yaghmazadeh et al.'s report demonstrates absence of a 'biophysical' non-thermal effect of RF energy on neural activity. On this account, these findings are not contradictory. The illustration in Fig. 2 summarizes these different effect mechanisms and the possibility of inducing non-thermal changes in the brain.

## V. A NEW WAVE OF EXPERIMENTAL STUDIES

One of the major drawbacks in the assessment of biological effects of RF energy radiation, that has played an important role in long-lasting uncertainties in the field, is the complexity of the required experimental evaluations. RF energy radiation interferes with metallic parts of the recording systems that are employed for measurement of biological variables, e.g., implanted electrodes that are used in electrophysiological recordings. Therefore, to ensure interference-free recordings, metal-free systems should only be applied. Recent technological advances in the development of optical methods for biological measurements have helped overcome this

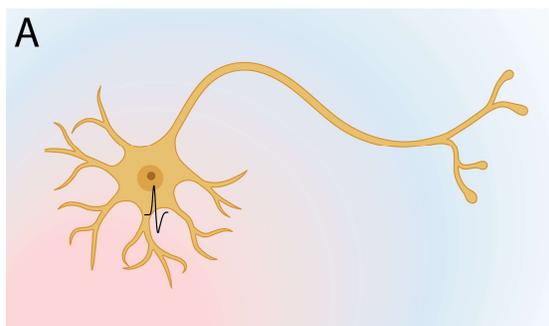 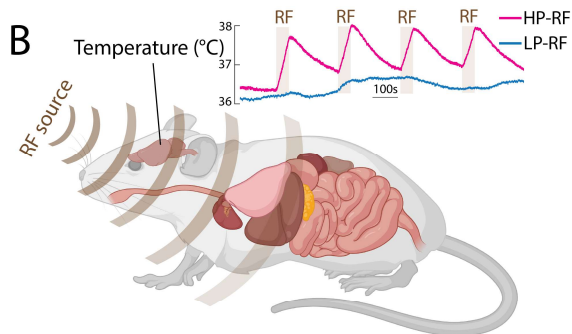

Biophysically *thermal* or *non-thermal*    Biologically *thermal* or *non-thermal*

Fig. 1. Thermal or Non-thermal categorization of effects can be done from the 'Biophysical' or the 'Biological' point of view. From the 'Biophysical' point of view, thermal effects are those whose mechanisms are thermally-mediated (A; a temperature gradient in the vacancy of the neuron can induce changes in its neural activity) and from the 'Biological' point of view, thermal effects are those that are accompanied with a temperature change in a biological meaningful manner (B; here when an RF source with sufficient power is used a temperature change in the animal's brain and/or body is observed, while a low-power source would not induce significant temperature changes (biologically non-thermal)).

challenge [5], [7]. Optical measurement methods mostly employ metal-free recording elements and have none or minimal interference with the RF field. Another issue to overcome is the interference of the RF energy radiation with the electronic components of the recording system. Even in optical measurements, the electrical parts of the recording system can be affected by RF interference and result in artifacts. Intensive care is needed to avoid such interferences (e.g. by using fiber-coupled solutions and putting electronic parts as far as possible from the RF source [7]).

## VI. THE HAVANA SYNDROME

Recently, the occurrence of the "Havana Syndrome" focused the collective attention towards possible significant effects of RF energy on the brain. Several American and Canadian diplomats across the globe reported hearing of sounds followed by some health effects with mostly neurological signature. It is generally recognized among the scientific community that the most likely cause of such effects, within the actual human technologies, is 'Directed RF energy' [8]. Dagro et al.'s paper introduces a possible mechanism in which a single extremely high-power short microwave pulse can induce lesion in brain tissue. However, the incident power levels to achieve such effects ($>10^6$ mW/cm$^2$) are too high to be caused by a distant source, which seems to be the case based on the descriptions of the events. Although the RF exposure paradigm in this study might not precisely explain the potential mechanism underlying the Havana syndrome, it is nevertheless very useful for understanding how RF pulse can induce thermos-mechanical forces in the brain. Hao et al.'s paper, on the other hand, introduced a paradigm in which repetitive sub-millisecond RF pulses are capable of inducing undesirable neurological effects at much lower power densities (200 mW/cm$^2$) that are possible to achieve from distant extremely high-power sources (which are currently used in military, or non-military HPM applications). Such pulse trains are thus the most likely mechanism underlying the Havana Syndrome.

## VII. THE NECESSITY TO UPDATE SAFETY GUIDELINES

The HPM technology has recently gained increasing interest for military and civil applications. HPM applications consist of high peak power bursts of narrowband RF energy in the frequency range of 1-100 GHz [9]. The recent reports featured in this piece which highlight the effects of pulsed RF energy on the brain raise serious questions about the safety of HPM applications. To date, the regulatory guidelines (such as those developed by the International Commission on Non-Ionizing Radiation Protection (ICNIRP) [10] and the Institute of Electrical and Electronics Engineers (IEEE) [11]) are focused on protection against dangers caused by thermal deposition in the biological tissue. As stated in both Dagro et al. and Hao et al.'s reports, high-power short RF pulses can lead to brain structural or functional damage without temperature increases beyond the biological range and in compliance with current safety regulations. This illustrates the necessity of the reevaluation of such regulatory guidelines to impose safe practices for application of high-power pulsed RF energy.

ACKNOWLEDGMENT

O.Y. thanks Nick Rommelfanger (Stanford University) for his helpful feedbacks on the manuscript.

AUTHOR CONTRIBUTIONS

O.Y. conceived and wrote this commentary piece.

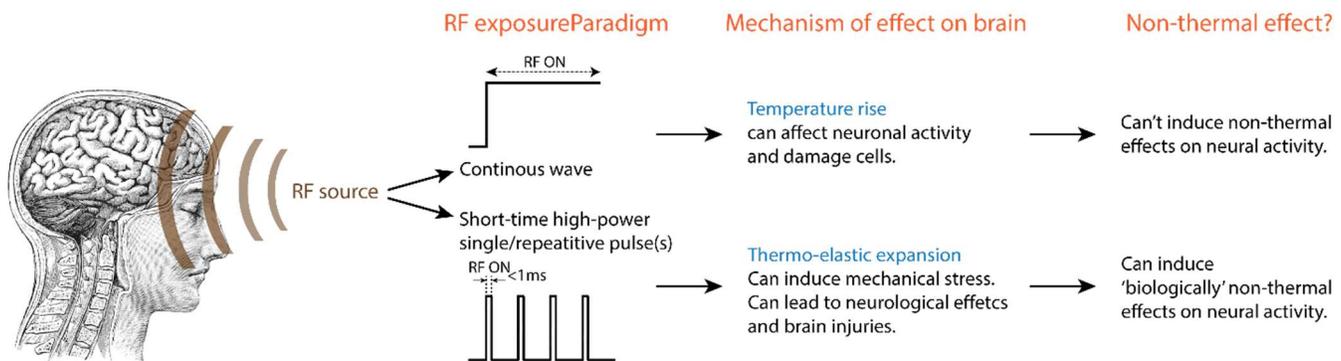

Fig. 2. Illustration summarizing the different mechanisms and consequences of CW RF and pulsed high-power RF.